\documentclass[aps,PRB,twocolumn,superscriptaddress,floats]{revtex4}
\usepackage{txfonts}
\usepackage{amssymb}
\usepackage{graphicx}

\begin{document}

\title{Angle-resolved photoemission spectroscopy study on iron-based superconductors}
\author{Z. R. Ye}
\author{Y. Zhang}
\author{B. P. Xie}\email{bpxie@fudan.edu.cn}
\author{D. L. Feng}\email{dlfeng@fudan.edu.cn}
\affiliation{State Key Laboratory of Surface Physics, Department of
Physics, and Advanced Materials Laboratory, Fudan University,
Shanghai 200433, People's Republic of China}

\begin{abstract}

Angle-resolved photoemission spectroscopy (ARPES) has played an important role in determining the band structure and the superconducting gap structure of iron-based superconductors. Here from the ARPES perspective, we briefly review the main results from our group in the recent years on the iron-based superconductors and their parent compounds, and depict our current understanding on the antiferromagnetism and superconductivity in these materials.

\end{abstract}

\maketitle

\section{Introduction}

Iron-based superconductors  belong to a new family of unconventional superconductors with the maximum superconducting transition temperature ($T_C$) up to 56~K in the bulk materials \cite{Hosono}. It provides a new route to realize and understand the high-$T_C$ superconductivity. So far, many series of iron-based superconductors have been discovered, which could be divided into iron-pnictides and iron-chalcogenides according to the anions \cite{Review}.  The undoped compounds of iron-based superconductors are usually in an antiferromagnetically ordered spin-density-wave (SDW) state. Through chemical substitution or physical pressure, the magnetic order is suppressed and the superconductivity emerges.

After five years' intensive research, many general consensuses have been reached on various critical issues of the physics of iron-based superconductors. During the process, as a powerful technique to study the electronic structure of solids,  angle-resolved photoemission spectroscopy (ARPES) plays an important role.  In this paper, we present a brief review of the ARPES studies on the iron-pnictides and iron-chalcogenides superconductors conducted by our group. We will discuss the multi-band and multi-orbital nature of the electronic structure, the mechanism of the structural and magnetic transitions, and the superconducting gap distributions etc. Meanwhile, we will discuss some ongoing issues which are still under debate,  such as the superconducting pairing symmetry in K$_x$Fe$_{2-y}$Se$_2$, nodal superconducting gap distribution, and the nematic transition prior to the AFM state.


\section{Angle-resolved photoemission spectroscopy}

\subsection{General description}

Based on the photoelectric effect, when a beam of monochromatized radiation supplied either by a gas-discharge lamp or a synchrotron beamline is shined on a sample, electrons are emitted and escape to the vacuum in all directions. An energy conservation during this photoemission process can be described by:
\[
E_{kin}  = h\nu  - |E_B|  - \varphi
\]
where $E_{kin}$ is the kinetic energy of the photoelectron, $h\nu$ is the photon energy,  $E_B$ is the electron binding energy, and $\varphi$ is the work function of the solid. Particularly, for the photoemission from solids with crystalline order, there is a momentum conservation between the in-plane photoelectron momentum ${\bf{p}}_\parallel$ and the crystal momentum of electron $\hbar {\bf{k}}_\parallel$, which can be described by:
\[
{\bf{p}}_\parallel   = \hbar {\bf{k}}_\parallel   = \sqrt {2mE_{kin} } \sin \theta
\]

Therefore, in order to obtain the binding energy and crystal momentum of the electron in the solid, we could measure the kinetic energy $E_{kin}$ of the photoelectron and its emission angle $\theta$ by using a hemisphere energy analyzer. The typical energy and angular resolutions are 5~meV and 0.3 degree respectively.

Due to the lack of translational symmetry along the sample surface normal,  the out-of-plane momentum ${\bf{k}}_\bot$ (or $k_z$) is not conserved during the photoemission process.  Such uncertainty in $k_z$ is less relevant in the case of low-dimensional systems, such as the  cuprates. However, it  is very important to estimate $k_z$ for studying the iron-based superconductors, whose electronic structure often has considerable variation along the  $k_z$ direction.  Fortunately in  ARPES experiments  at fixed ${\bf{k}}_\parallel$,
the  band dispersion along the ${\bf{k}}_\bot$ direction could still be measured  by varying the photon energy, since it  changes $E_{kin}$. The ${\bf{k}}_\bot$  could be approximately calculated by  using ``inner potential'' $V_0$  as the following:
\[
{\bf{k}}_\bot = \hbar^{-1} \sqrt{2m(E_{kin}\cos^2\theta + V_0)}
\]

In practice,  $V_0$  is chosen so that  the  periodicity of the Brillouin zone in the ${\bf{k}}_\bot$  direction is reproduced correctly in the measured dispersion.
Therefore, conducting the photon energy dependent ARPES  is an effective way to  probe the      electronic structure in the three-dimensional (3D) Brillouin zone, which is crucial for the study of iron-based superconductors.

\subsection{Polarization dependent ARPES}

\begin{figure}[t]
\includegraphics[width=8.7cm]{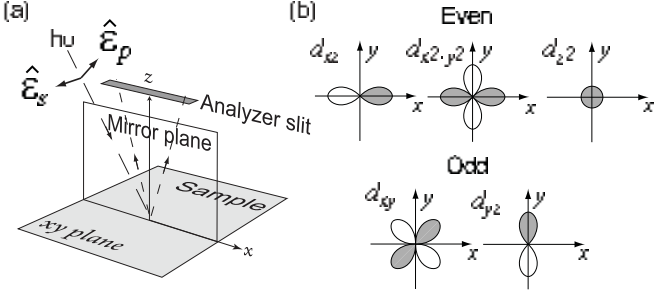}
\caption{(a) Experimental setup for polarization-dependent ARPES. For
the $p$ (or $s$) experimental geometry, the electric field direction
of the incident photons $\bf{\hat{\varepsilon}_{p}}$ (or
$\bf{\hat{\varepsilon}_{s}}$)  is parallel (or perpendicular) to the
mirror plane defined by the analyzer slit and the sample surface
normal. (b)  Illustration of the spatial symmetry of the
$3d$ orbitals with respect to the $xz$ plane. } \label{orbital1}
\end{figure}

The polarization-sensitivity of orbitals in ARPES is its another advantage in studying the iron-based superconductors  \cite{ReviewArpes, YZhangBaCo}.  The photoemission intensity is proportional to the matrix element of the photoemission process ${I_0}(\bf{k}, v, \bf{A})$ $ \propto$ ${|M_{f,i}^{\bf{k}}|^2}$, which can be described by ${|M_{f,i}^{\bf{k}}|^2\propto{\rm{|}}\langle \phi _f^{\bf{k}}|\bf{\hat{\varepsilon}}\cdot{\bf{r}}|\phi _i^{\bf{k}} \rangle |^2}$, where
$\bf{\hat{\varepsilon}}$ is the unit vector of the electric field of the light,  and ${\phi _i^{\bf{k}}}$ (${\phi _f^{\bf{k}}}$) is the initial-state (final-state) wave function.  In order to have nonvanishing photoemission intensity, the whole integrand in the overlap integral must be an even function  under reflection with respect to the mirror plane defined by the analyzer slit and the sample surface normal [Fig.~\ref{orbital1}(a)]. Because odd-parity final states would be zero everywhere on the mirror plane and therefore also at the detector, the final state ${\phi _f^{\bf{k}}}$ must be even.  In particular, for high kinetic-energy photoelectrons, the final-state wave function ${\phi _f^{\bf{k}}}$ can be approximated by an even-parity plane-wave state ${e^{i{\bf{k}\cdot\bf{r}}}}$ with $\bf{k}$ in the mirror plane. In turn, this implies that $\bf{\hat{\varepsilon}}\cdot{\bf{r}}|\phi _i^{\bf{k}} \rangle$ must be even. For the $p$ (or $s$) experimental geometry, where the electric field direction of the incident photons $\bf{\hat{\varepsilon}_{p}}$ (or $\bf{\hat{\varepsilon}_{s}}$) is parallel (or perpendicular) to the mirror plane [Fig.~\ref{orbital1}(a)],  $\bf{\hat{\varepsilon}}\cdot{\bf{r}}$ will be even (or odd). Therefore,  only the even (or odd) parity initial states ${\phi _i^{\bf{k}}}$ will be detected \cite{ReviewArpes}.

Considering the spatial symmetry of the $3d$ orbitals, when the analyzer slit is along the high-symmetry direction of the sample, the photoemission signal of certain orbital would appear or disappear by specifying the polarization directions. For example, with the analyzer slit in the $xz$ plane [Fig.~\ref{orbital1}(a)], the even orbitals ($d_{xz}$, $d_{z^2}$, and $d_{x^2-y^2}$) and the odd orbitals ($d_{xy}$ and $d_{yz}$) could be only observed in the $p$ and $s$ geometries, respectively [Fig.~\ref{orbital1}(b)].

\begin{figure}[t]
\includegraphics[width=8.7cm]{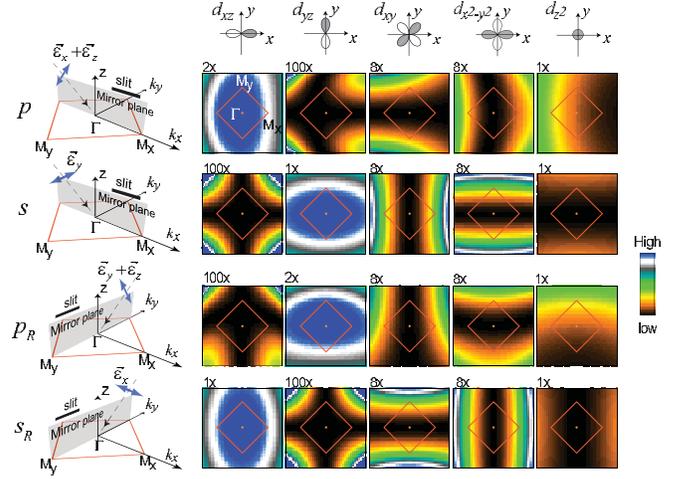}
\caption{The experimental setup and the corresponding simulated matrix element  for the $3d$ orbitals. The mirror planes are defined by the analyzer slit and the sample surface normal. The two-dimensional plot of the Brillouin zone is illustrated by red solid squares.  Note that the photoemission cross-sections are amplified by a factor shown at the up-left corner of each panel. Thus, all the panels could be shown in the same color scale. There is a minor asymmetry in certain distributions caused by the out-of-plane component of the polarization.
} \label{orbital2}
\end{figure}

One could further change the mirror plane by rotating the azimuthal angle of the sample. Four experimental geometries could be achieved. The matrix element distributions for the five orbitals were calculated in four geometries. As shown in Fig.~\ref{orbital2}, the matrix element distributions of the $d_{xz}$ and $d_{yz}$ orbitals exhibit a strong polarization dependence throughout the first Brillouin zone, reflecting the opposite symmetry of these two orbitals. The matrix element distribution of $d_{xy}$ ($d_{x^2-y^2}$) is suppressed along the direction parallel (perpendicular) to the in-plane component of the polarization. For the $d_{z^2}$ orbital which is sensitive to  the out-of-plane component of the polarization, its intensity is much stronger in the $p$ experimental geometry than in the $s$ experimental geometry. Note that, we use the atomic wave function of $3d$ electrons as the initial wave function for simplicity. More strictly speaking,  the Bloch wave function need to be used in the analysis \cite{Bruet}. 

Therefore, by comparing the experimental results with the calculations, we could determine the orbital characters of individual bands and further study the role of the orbital degree of freedom in iron-based superconductors.

\section{Iron-pnictides}

\subsection{Parent compounds}

\begin{figure}[t]
\includegraphics[width=8.7cm]{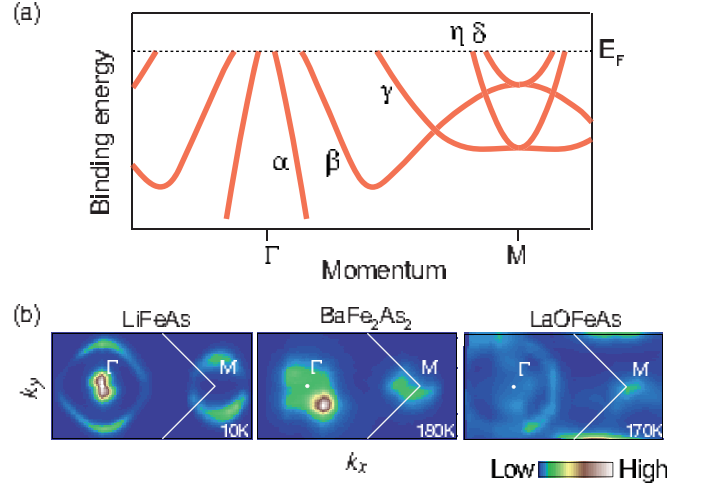}
\caption{(a) Cartoon of the band structure in iron-pnictides. (b) Fermi surface mappings of LiFeAs, BaFe$_2$As$_2$,  and LaOFeAs respectively. } \label{FS}
\end{figure}

\begin{figure}[t]
\includegraphics[width=8.7cm]{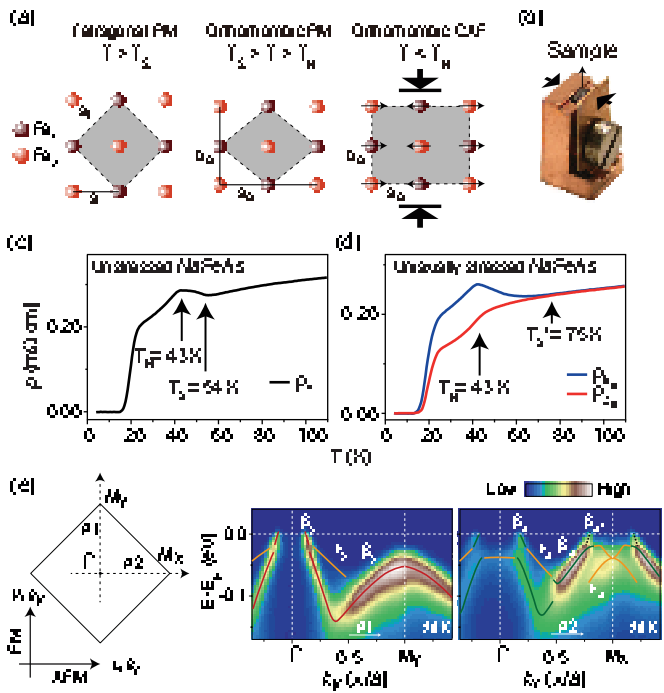}
\caption{(a) Cartoon of the lattice and spin structure in tetragonal paramagnetic (PM), orthorhombic PM, and orthorhombic CAF state for iron-pnictides. The $x$ and $y$ axes are defined along the iron-iron directions. The black arrows show the direction of the uniaxial pressure applied in the mechanical detwinning process. (b) Photograph of the device used to detwin the samples in our experiments. (c) and (d) The temperature dependent resistivity of unstressed and uniaxially stressed NaFeAs, respectively. (e) The Photoemission intensities taken at 20~K in uniaxially stressed NaFeAs along $\Gamma$-$M_y$ and $\Gamma$-$M_x$ directions respectively. Reprinted with permission from \cite{NaFeAsPRB}, copyright 2012 by the American Physical Society. } \label{SDW1}
\end{figure}

Unlike the cuprates, the parent compounds of iron-pnictides are metals, instead of insulators. Early band calculations showed that the low-lying electronic structure is dominated by the Fe $3d$ electrons \cite{Theory1, Theory2}. There are three hole-like bands around $\Gamma$ and two electron-like bands around M [Fig.~\ref{FS}(a)], exhibiting a semi-metal behavior. Our ARPES measurements on several parent compounds have confirmed this scenario \cite{lxyang, lxyangPRB, Hecheng}. As shown in Fig.~\ref{FS}(b), the photoemission intensity around $\Gamma$ is contributed by hole pockets, while that around $M$ is contributed by electron pockets. Note that, in  LiFeAs and BaFe$_2$As$_2$, the sizes of the electron and hole pockets are comparable, while the huge hole pocket in LaOFeAs is due to the charge redistribution on the surface of the 1111 compounds \cite{lxyangPRB}. Polarization dependent ARPES studies and band calculations further showed that the hole and electron pockets involve all five Fe $3d$ orbitals \cite{YZhangBaCo}. Therefore, in contrast to the single band Fermi surface of cuprates, the electronic structure of iron-based superconductors exhibits multi-band and multi-orbital nature.

Besides the similar electronic structure, the parent compounds of iron-pnictides share a common SDW or collinear antiferromagnetic (CAF)  ground state \cite{Neutron1, Neutron2}, which is characterized by a ferromagnetic (FM) spin alignment along one direction in the two-dimensional rectangular lattice formed by iron sites, and an antiferromagnetic (AFM) spin alignment along the perpendicular direction [Fig.~\ref{SDW1}(a)]. Intriguingly, the development of the SDW order is always accompanied by a tetragonal-to-orthorhombic structural phase transition. The transition temperature ($T_S$) at which the lattice distortion takes place either precedes or coincides with the  Neel transition temperature ($T_N$) \cite{Neutron1, Neutron2}.

Early ARPES measurements are complicated due to the twinning effect of the sample \cite{lxyang, Hecheng}, since the $C4$ rotational symmetry is broken via the structural and magnetic transitions. Therefore, in order to investigate the low-temperature electronic state, we applied a uniaxial pressure on the sample to overcome the twinning effect [Fig.~\ref{SDW1}(b)]. After detwinning, the in-plane anisotropy of the resistivity emerges at about 75~K corresponding to the nematic transition temperature [Fig.~\ref{SDW1}(d)], much higher than $T_S$ and $T_N$ in the unstressed sample [Fig.~\ref{SDW1}(c)]. Meanwhile, we have successfully observed a nematic electronic structure with $C2$ rotational symmetry in the SDW state by ARPES \cite{NaFeAsPRB}. As shown in Fig.~\ref{SDW1}(e), the band structures along $\Gamma$-$M_x$ and $\Gamma$-$M_y$ direction are different. Note that, the electronic structure in the SDW state is characterized by a strong band reconstruction rather than a Fermi-surface-nesting gap.

\begin{figure}[t]
\includegraphics[width=8.7cm]{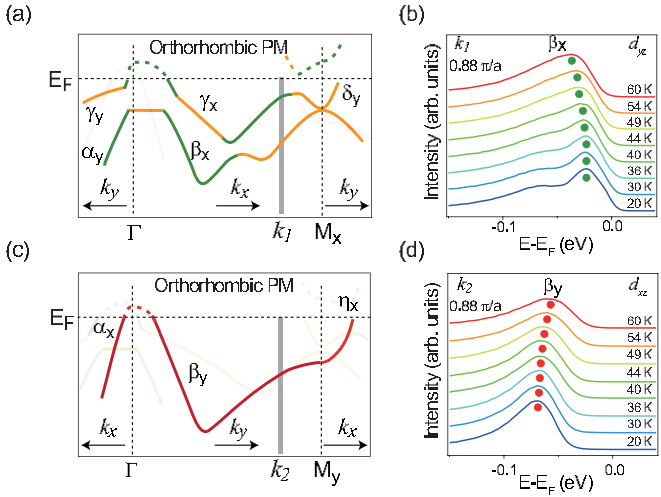}
\caption{ (a) The band structure in the orthorhombic PM state,  where only $d_{yz}$- and $d_{xy}$-dominated bands are highlighted. (b) The temperature dependence of the EDCs at $k_1$ as indicated by the gray line in panel (a). (c) and (d) are the same as panels (a) and (b), respectively, but for the $d_{xz}$-dominated bands. Reprinted with permission from \cite{NaFeAsPRB}, copyright 2012 by the American Physical Society.} \label{SDW2}
\end{figure}

\begin{figure}[t]
\includegraphics[width=8.7cm]{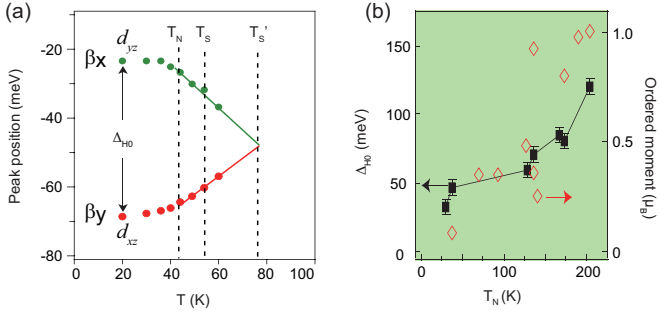}
\caption{  (a) The peak positions of the $\beta_x$ and $\beta_y$ bands as functions of temperature. We define the maximal observable separation between $\beta_x$ and $\beta_y$ at the same mometum value (\textit{i.e.} $|k_x|=|k_y|$) near $M_x$ and $M_y$ respectively as $\Delta_{H}$.  $\Delta_{H}$ is a function of temperature, and its low temperature saturated value is defined as $\Delta_{H0}$. (b) $\Delta_{H0}$ obtained from our ARPES data (including both the published and the unpublished), and the low temperature ordered moment measured by neutron scattering in various iron pnictides are plotted as a function of the Neel temperature, and the data are also tabularized in the Table 1. Panel (a) is reprinted with permission from \cite{NaFeAsPRB}, copyright 2012 by the American Physical Society. Panel (b) is reprinted with permission from \cite{JJ}.} \label{SDW3}
\end{figure}

The energy separation of $\beta_x$ and $\beta_y$ near the $M_x$ or $M_y$ point could be viewed as a parameter to describe the reconstruction of the electronic structure.  Such a separation begins at the nematic transition temperature and almost saturates at $T_N$ [Figs.~\ref{SDW2} and \ref{SDW3}(a)]. The smooth evolution of the electronic structure reconstruction across $T_S$ into the SDW state [Figs.~\ref{SDW2}(b) and \ref{SDW2}(d)] indicates that both the magnetic and structural transitions share the same driving force \cite{NaFeAsPRB}. We further summarized the energy separation between $\beta_x$ and $\beta_y$ measured in various iron-based compounds [Fig.~\ref{SDW3}(b)]. Their low-temperature saturated values ($\Delta_{H0}$) roughly show a monotonic correspondence with  $T_N$'s \cite{JJ}. Together, the ordered moments measured by neutron scattering for various compounds are plotted against $T_N$. The similar trends in both quantities suggest that the band sepration is induced by the Hund's rule coupling between the itinerant electrons and the local moments.

With polarization dependent ARPES, we found that the band reconstruction primarily involves the $d_{xy}$- and $d_{yz}$-dominated bands. These bands strongly hybridize with each other, inducing a band splitting, while the $d_{xz}$-dominated bands only exhibit an energy shift without any reconstruction \cite{NaFeAsPRB}. As a result, the orbital weight redistribution of $d_{yz}$ and $d_{xy}$ opens a partial gap near the Fermi energy ($E_F$). On the contrary, the total occupation of $d_{xz}$ is almost invariant \cite{NaFeAsPRB}. Theoretically, either  exchange interactions among spins or ferro-orbital ordering between $d_{xz}$ and $d_{yz}$ orbitals has been suggested to be the cause of the SDW order \cite{Theory3, Theory4}. Our data suggest that the  fluctuations of the spin order at high temperatures and the local Hund's rule coupling  drive the nematic electronic structure in iron-pnictides, while the weak ferro-orbital ordering (or orbital occupation redistribution) between the $d_{xz}$ and $d_{yz}$ orbitals may be just a consequence or  a secondary driving force here \cite{NaFeAsPRB}.

\begin{table}[!htbp]
\caption{Data of $T_N$(K),  $\Delta_{H0}$(meV), $S$($\mu_B$) in Fig. 6.}
\begin{tabular}{|l|l|l|l|}
\hline
Compound & $T_N$(K) & \hspace{10pt} $\Delta_{H0}$ (meV) \hspace{10pt}  & \hspace{10pt} $S$($\mu_B$) \hspace{10pt}\\
\hline
SrFe$_2$As$_2$ & 205 & 120 (Ref.~\onlinecite{SrK}) & 1.01(3) (Ref.~\onlinecite{SrFeAs})  \\
EuFe$_2$As$_2$ & 190 & - & 0.98(8) (Ref.~\onlinecite{EuFeAs})  \\
CaFe$_2$As$_2$ & 173 & 80 & 0.80(5) (Ref.~\onlinecite{CaFeAs}) \\
Sr$_0$$_.$$_9$K$_0$$_.$$_1$Fe$_2$As$_2$ & 168 & 85  (Ref.~\onlinecite{SrK})& - \\
NdOFeAs & 141 & - & 0.25(7) (Ref.~\onlinecite{NdOFeAs})  \\
LaOFeAs & 137 & - & 0.36  (Ref.~\onlinecite{PCDai})  \\
BaFe$_2$As$_2$ & 136 & 70  (Ref.~\onlinecite{MYi}) & 0.93(6) (Ref.~\onlinecite{BaFeAs})  \\
Sr$_0$$_.$$_8$$_2$K$_0$$_.$$_1$$_8$Fe$_2$As$_2$ & 129 & 60  (Ref.~\onlinecite{SrK}) & -  \\
PrOFeAs & 127 & - & 0.48(9) (Ref.~\onlinecite{PrOFeAs})  \\
BaFe$_0$$_.$$_9$$_5$Co$_0$$_.$$_0$$_5$As$_2$ & 93 & - & 0.35  (Ref.~\onlinecite{BaCo})  \\
Ba$_1$$_-$$_x$K$_x$Fe$_2$As$_2$ & 70 & - & 0.35  (Ref.~\onlinecite{BaK})  \\
NaFeAs & 37 & 46  (Ref.~\onlinecite{NaFeAsPRB}) & 0.09(4) (Ref.~\onlinecite{NaFeAs})  \\
NaFe$_0$$_.$$_9$$_8$$_7$$_5$Co$_0$$_.$$_0$$_1$$_7$$_5$As & 30 & 32  (Ref.~\onlinecite{Ge}) & -\\
\hline
\end{tabular}
\end{table}


\subsection{Superconducting iron pnictides}

\subsubsection{Coexistence of SDW and supercondcutivity}

The superconductivity in iron pnictides could be induced, when the SDW order is suppressed by chemical substitution of the parent compounds in various ways \cite{Review}.  For example, the SDW is suppressed in Sr$_{1-x}$K$_{x}$Fe$_2$As$_2$ by  K doping.  As shown in Fig.~\ref{SDW3}(b),  both $T_N$ and the band separation $\Delta_{H0}$,  which is an electronic hallmark of the SDW state, decrease with increasing K concentration in underdoped Sr$_{1-x}$K$_{x}$Fe$_2$As$_2$  \cite{SrK}. Interestingly, in the superconducting sample Sr$_{0.82}$K$_{0.18}$Fe$_2$As$_2$, one could still observe a clearly band separation below 130~K \cite{SrK}.   This shows  that the superconductivity and the SDW might coexist in the underdoped regime of the phase diagram. In fact, such a coexistence of superconductivity and SDW has been found to be quite generic in various iron pnictides.

\begin{figure}[t]
\includegraphics[width=8.7cm]{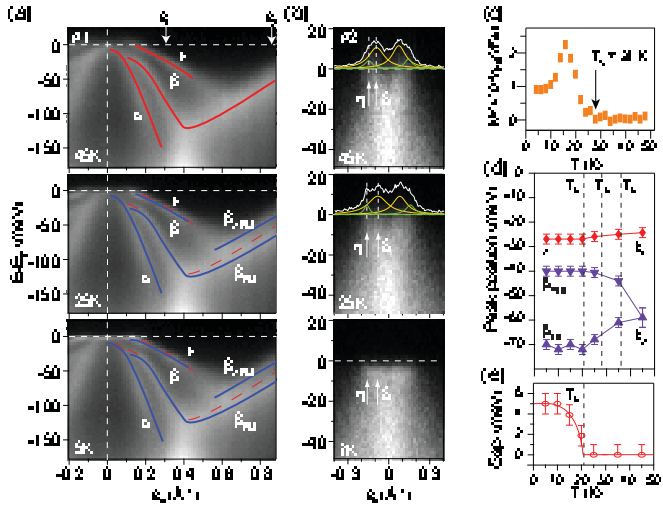}
\caption{(a) The band structure of NaFe$_{0.9825}$Co$_{0.0175}$As at 45, 25, and 5 K respectively along  $\Gamma$~-~M direction.
The dashed lines in the lower panels are the band dispersion at 45~K for comparison purpose. (b) Temperature dependence of the band structure around the zone
corner. The MDCs at E$_{F}$ are plotted on the 25 and 45 K data. Each MDC was fitted to four Lorentzians (overlaid yellow and green lines).
(c) Temperature dependence of the magnetic order parameter at Q= (1, 0, 1.5) for NaFe$_{0.9825}$Co$_{0.0175}$As measured by neutron scattering. (d)  The temperature dependence
of the  peak positions of the EDCs taken at k$_1$ and k$_2$ as marked in panel (a). (e) The temperature dependence of the superconducting gap of $\gamma$. The gap size is estimated through an empirical fit as described in detail
in Ref.~\cite{Zhangnode}. Reprinted with permission from \cite{Ge}, copyright 2013 by the American Physical Society. } \label{coexist2}
\end{figure}

\begin{figure}[t]
\includegraphics[width=8.7cm]{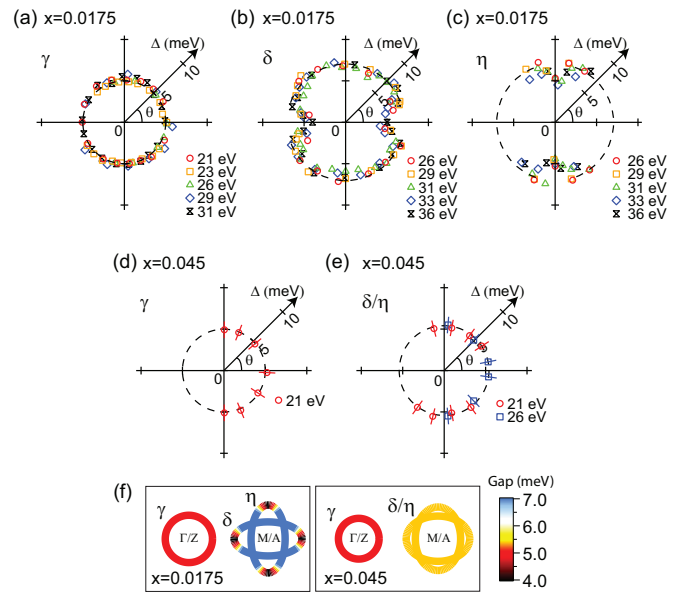}
\caption{Polar plots of the superconducting gap for the (a) $\gamma$, (b)
$\delta$ and (c) $\eta$ Fermi surfaces of NaFe$_{0.9825}$Co$_{0.0175}$As,
respectively.  The error bar for the gaps is
$\pm$1~meV based on the fitting. Polar plots of the superconducting gap of NaFe$_{0.955}$Co$_{0.045}$As for the (d) $\gamma$,  and (e) $\delta/\eta$
Fermi surfaces respectively. (f) False-color plots of the gap distribution
on the Fermi surfaces of NaFe$_{0.9825}$Co$_{0.0175}$As and NaFe$_{0.955}$Co$_{0.045}$As. Reprinted with permission from \cite{Ge}, copyright 2013 by the American Physical Society.  } \label{coexist3}
\end{figure}

The SDW/superconductivity  coexisting phase in the iron pnictides represents a novel ground state, and its properties may shed light on the superconducting mechanism. Theoretically, theories based on $s{++}$ pairing symmetry suggest that
there must be nodes in the superconducting  gap in this regime \cite{theoryMazin}
and the coexisting SDW and superconducting phases cannot be microscopic \cite{NeutronSc&Theory}.
On the other hand, theories based on $s\pm$ pairing symmetry suggest
nodeless superconducting gap in the presence of weak magnetic order; moreover,
the coexistence may cause angular variation of the superconducting gap, and even give
rise to nodes in the limit of strong antiferromagnetic ordering
\cite{theoryMazin,theoryChubukov}.

We have conducted ARPES experiments on the 1.75\% Co doped NaFeAs, which is in the coexisting regime \cite{Ge}. The neutron scattering and transport measurements confirmed the coexistence of SDW and superconductivity with $T_N$~=~28~K and $T_c$~=~20.5~K. As shown in Fig.~\ref{coexist2}, the band structure reconstruction corresponding to the SDW formation and the superconducting gap could be observed on the same $\gamma$, $\eta$ and $\delta$ bands, which is a direct evidence for the intrinsic coexistence of the two orders. 
Since the reconstruction of electronic structure in the SDW state does not necessarily open an full energy gap at $E_F$, which leaves room for superconductivity.
More intriguingly, the superconducting gap distribution
is found nodeless on all Fermi surface sheets: it is isotropic on
the hole pocket, but it is highly anisotropic on the electron pockets [Figs.~\ref{coexist3}(a)$\sim$ \ref{coexist3}(c)]. However, for the comparison measurements on an SDW-free NaFe$_{0.955}$Co$_{0.045}$As sample ($T_{c}=20$~K), in-plane gap distributions on all Fermi surface sheets show isotropic character [Figs.~\ref{coexist3}(d) and \ref{coexist3}(e)]. Since NaFe$_{0.9825}$Co$_{0.0175}$As and
NaFe$_{0.955}$Co$_{0.045}$As have similar Fermi surfaces, orbital characters and
interaction parameters, the highly anisotropic gap distribution on the electron pockets of
NaFe$_{0.9825}$Co$_{0.0175}$As is most likely a direct consequence
of the coexisting SDW. Our data thus could be viewed as a positive support for the $s\pm$ pairing symmetry in iron pnictides.

\subsubsection{The nodal superconducting gap }

\begin{figure}[t]
\includegraphics[width=8.7cm]{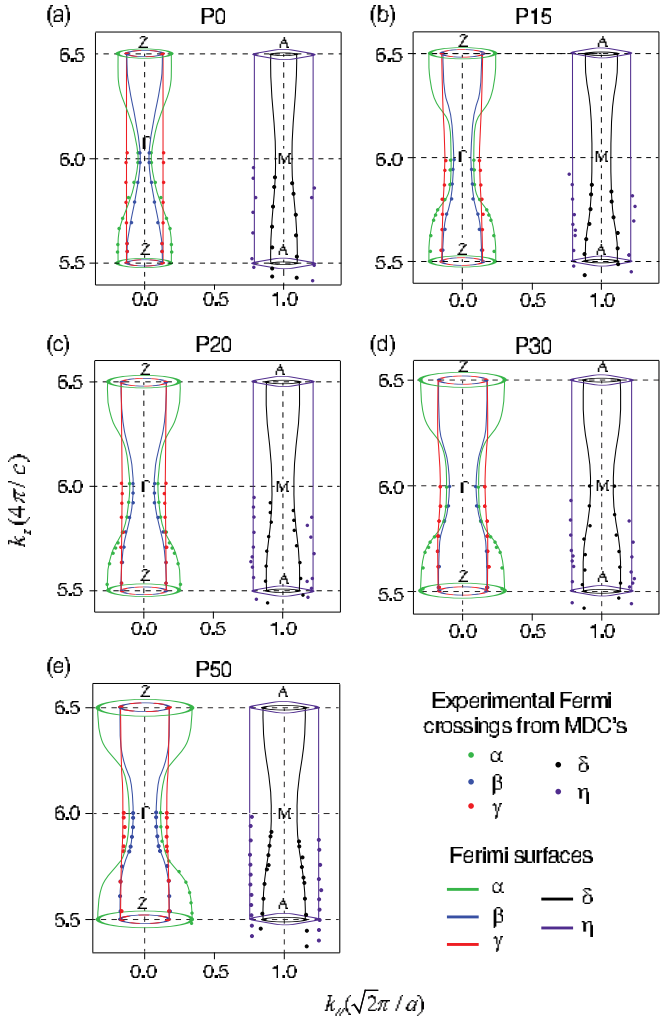}
\caption{Doping dependence of the experimental  Fermi surface cross-sections in
Z-$\Gamma$-M-A plane for BaFe$_2$(As$_{1-x}$P$_x$)$_2$.  Reprinted with permission from \cite{zirongye}, copyright 2012 by the American Physical Society.} \label{FS3D}
\end{figure}

Pairing symmetry is a pivotal characteristic for a superconductor. In the conventional BCS superconductors, the formation of Cooper pairs is due to the attractive interaction between electrons mediated by the electron-phonon interaction. Such pairing interaction results in an isotropic $s$-wave pairing symmetry. However, for  cuprates, since the Coulomb repulsive interaction between electrons is rather strong,   an $d$-wave pairing symmetry is favored energetically.  The nodal gap structure measured by ARPES has played a critical role in establishing the $d$-wave pairing picture there. 

For iron-based superconductors, the situation is  more complicated.  There are both nodal and nodeless (gap)  iron-based superconductors. The nodeless gap distributions, which could be attributed to  $s$-wave or s$\pm$ pairing symmetry, have been directly observed by ARPES and other techniques in Ba$_{1-x}$K$_{x}$Fe$_2$As$_2$, Ba(Fe$_{1-x}$Co$_x$)$_2$As$_2$, and so on \cite{Gap1, Gap2}. However, the signatures of nodal superconducting gap have been reported in LaOFeP, LiFeP, KFe$_2$As$_2$, BaFe$_2$(As$_{1-x}$P$_x$)$_2$, and BaFe$_{2-x}$Ru$_x$As$_2$ by thermal conductivity, penetration depth, nuclear magnetic resonance, and scanning tunneling microscope (STM) measurements \cite{linenode7, linenode1, linenode2, linenode3, linenode4, linenode5, linenode6}.

Many theories have been proposed to understand the nodal behavior in iron pnictides. Particularly, it is predicted that the $d_{xy}$-based band would move to higher binding energy with increasing P doping in BaFe$_2$(As$_{1-x}$P$_x$)$_2$, and nodes would appear on the electron pockets when the $d_{xy}$ hole Fermi pocket disappears. Figure \ref{FS3D} shows the experimental doping dependence of the three-dimensional Fermi surfaces in BaFe$_2$(As$_{1-x}$P$_{x}$)$_2$. It is found that the $d_{xy}$-originated $\gamma$ hole pocket is always present  for all dopings, which thus disproves the theories that explain the nodal gap based on the vanishing $d_{xy}$ hole pocket \cite{zirongye}.

\begin{figure}[t]
\includegraphics[width=8cm]{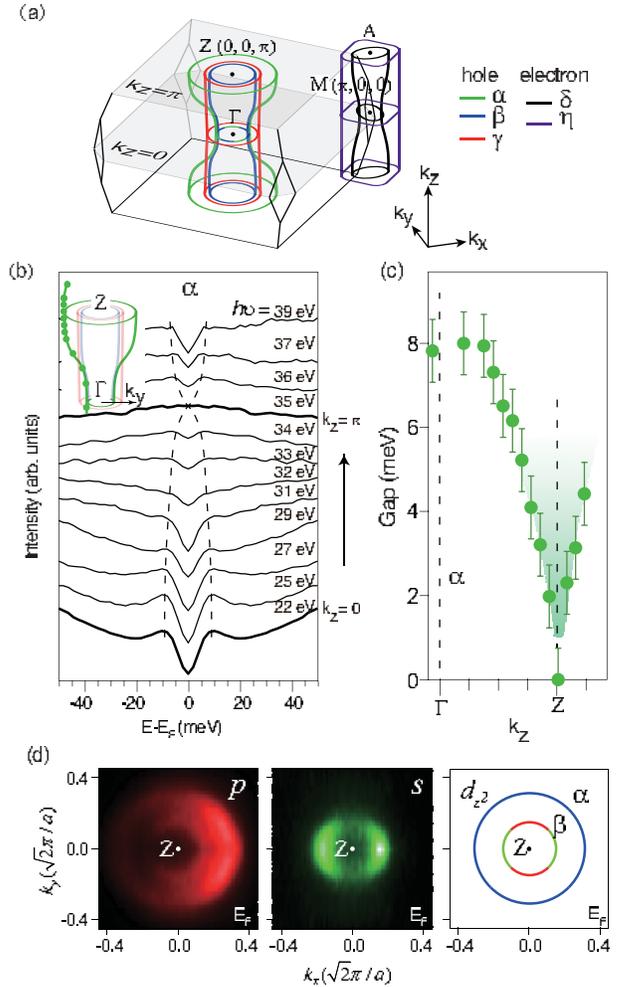}
\caption{(a) The three-dimensional Fermi surfaces of BaFe$_{2}$(As$_{0.7}$P$_{0.3}$)$_2$.  The two-iron unit cell is implemented here, with the Fe~-~Fe direction as the $k_x$ direction. The electron Fermi surfaces are only illustrated at one corner of the Brillouin zone for simplicity.  (b)  $k_z$ dependence of the symmetrized spectra
measured on the $\alpha$ hole Fermi surface. The symmetrized spectra near $k_z~=~0$ and $k_z~=~\pi$
are shown in thicker lines. The dashed line is a guide to the eyes for  the variation of the superconducting gap at different $k_z$'s. (c) The superconducting gap on the $\alpha$ Fermi surface with respect to $k_z$. (d) The photoemission intensity maps at $E_F$ for BaFe$_2$(As$_{0.7}$P$_{0.3}$)$_2$ taken
with 100~eV photons around the Z point at 40~K. The left panel shows data taken in the p-polarization, which exhibits the outer ring contributed by the $\alpha$ band. This outer ring is absent in the data shown in the middle panel taken in the s-polarization. Such a polarization dependence shows that the $\alpha$ band is mainly made of the $d_{z^2}$ orbital near Z as   summarized in the right  panel. Reprinted with permission from \cite{Zhangnode}, copyright 2012 by Nature Physics.} \label{BP1}
\end{figure}

\begin{figure}[t]
\includegraphics[width=7.5cm]{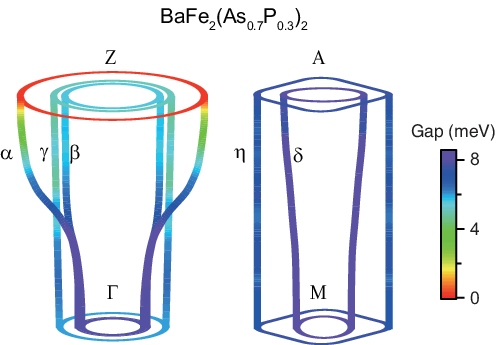}
\caption{The gap distribution on the Fermi surfaces of BaFe$_{2}$(As$_{0.7}$P$_{0.3}$)$_2$. Reprinted with permission from \cite{Zhangnode}, copyright 2012 by Nature Physics.} \label{BP2}
\end{figure}

The most straightforward way to understand the nodal behavior or the pairing symmetry is to determine the momentum distribution of the superconducting gap.  BaFe$_2$(As$_{0.7}$P$_{0.3}$)$_2$  is a typical iron pnictide with nodal superconducting gap. As shown in  Fig.~\ref{BP1}(a), it  has three hole Fermi surface sheets  ($\alpha$, $\beta$ and $\gamma$) surrounding the central $\Gamma$~-~Z axis of the Brillouin zone, and two electron Fermi surface sheets ($\eta$ and $\delta$) around the zone corner. 
Detailed survey on the electron Fermi surface sheets exhibited a nodeless superconducting gap with little $k_z$ dependence. However, for the $\alpha$  hole Fermi surface, the experimental data clearly showed a zero superconducting gap or nodes located around the Z point [Figs.~\ref{BP1}(b) and \ref{BP1}(c)] \cite{Zhangnode}.

The gap distribution of BaFe$_2$(As$_{0.7}$P$_{0.3}$)$_2$ is summarized in Fig.~\ref{BP2}. Such a horizontal line-node distribution immediately rules out the $d$-wave pairing symmetry, which would have given four vertical line nodes in the diagonal directions ($\theta$\,=\,$\pm 45^{\circ}$, $\pm 135^{\circ}$), as in the cuprates. For the $s$-wave pairing symmetry, the horizontal ring node around Z is not enforced by symmetry, as it is fully symmetric with respect to the point group. Therefore, the nodal ring is an ``accidental'' one, which is probably induced by the strong three-dimensional nature of the $\alpha$ band \cite{Kuroki, HuJP}, for example its sizable d$_{z^2}$ orbital character near Z [Fig.~\ref{BP1}(d)]. This also explains why the gap is nodal for certain compounds and nodeless for some other compounds.

\begin{figure}[t]
\includegraphics[width=8.7cm]{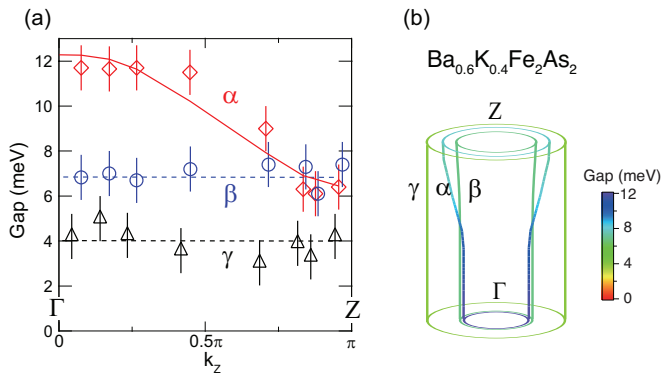}
\caption{(a) and (b),  $k_z$ dependence of the superconducting gaps around the zone center.  Reprinted with permission from \cite{YanBaK}, copyright 2010 by the American Physical Society. } \label{BK1}
\end{figure}

Furthermore, the strong $k_z$ dependence of the superconducting gap was also observed in Ba$_{0.6}$K$_{0.4}$Fe$_{2}$As$_2$ \cite{YanBaK}. As shown in Fig.~\ref{BK1}(a),  the superconducting gap on the $\alpha$ Fermi surface decreases significantly from $\Gamma$ to Z, while those on the $\beta$ and $\gamma$ Fermi surfaces are relatively unchanged. Meanwhile, the $k_z$ dispersion of the $\alpha$ band is much stronger than $\beta$ and $\gamma$ [Fig.~\ref{BK1}(b)].  The strong out-of-plane and orbital symmetry
dependence of the superconducting gap in  Ba$_{0.6}$K$_{0.4}$Fe$_{2}$As$_2$  and the horizontal nodal gap structure in BaFe$_2$(As$_{0.7}$P$_{0.3}$)$_2$ all indicate that the 3D electronic structure and multi-orbital nature play important roles in inducing the anisotropic or nodal gap structure in iron-based superconductors.

\subsubsection{Correlations between the Fermi surface topology and superconductivity in iron pnictides}

\begin{figure}[t]
\includegraphics[width=8.7cm]{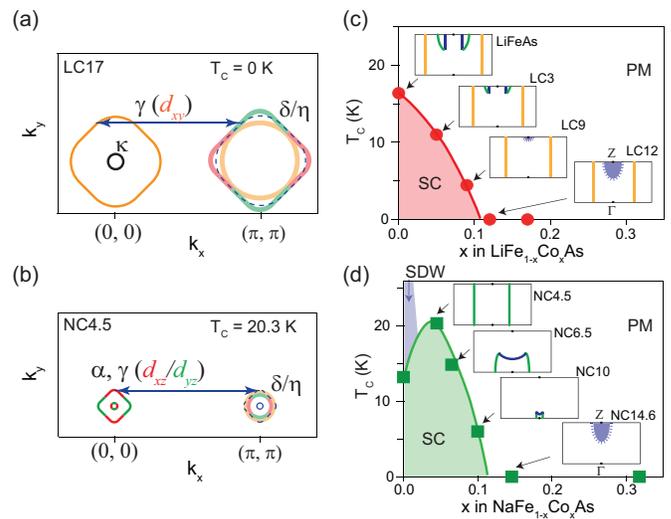}
\caption{(a) and (b), Illustration of the Fermi surface nesting condition in
LC17 and NC4.5 (named by their dopant percentages), respectively. (c) and (d),  The phase diagram and corresponding Fermi surface topology near zone center for LiFe$_{1-x}$Co$_x$As  and NaFe$_{1-x}$Co$_x$As, respectively. The solid lines represent hole Fermi surface sheets, while dashed ones with blue area inside represent electron Fermi surface sheets.  Reprinted with permission from \cite{zirong}. } \label{111}
\end{figure}

The correlation between the superconductivity in iron pnictides and their Fermi surface topology is elusive and controversial so far. Experimentally, at first,  the nesting between any zone center hole Fermi surface and any zone corner electron Fermi surface was suggested to be important for the superconductivity. Later on,  it was pointed out  in BaFe$_{2-x}$Co$_x$As$_2$ and LiFeAs that Fermi surface nesting is unnecessary, while the presence of central hole pockets or Van Hove singularity are more important \cite{HDingBaK, Kaminski, Borisenko}. Theoretically, a majority of  studies  indicate that the inter-pocket nesting would significantly enhance the superconductivity, while some predicted that the $T_C$ peaks near a Lifshitz transition  \cite{ KurokiT, Mazin, shape}. In general, it appears that these contradicting correlations between Fermi surface topology and superconductivity  are all partially supported by different experiments and theories, and there are always counter examples. This situation can be clarified when one realize that the correlation between Fermi surface topology and superconductivity is orbital-selective \cite{zirong}. For  both LiFe$_{1-x}$Co$_x$As and NaFe$_{1-x}$Co$_x$As series, it has been shown that the $T_C$ is maximized only by the perfect nesting  between $d_{xz}$/$d_{yz}$-originated Fermi surfaces [Figs.~\ref{111}(a) and \ref{111}(b)], while the superconductivity  diminishes quickly after the central $d_{xz}$/$d_{yz}$ hole Fermi surfaces disappear with electron doping [Figs.~\ref{111} (c) and \ref{111}(d)].  The orbital selective nature of these correlations clarify most previous controversies discussed above.



\section{Iron-chalcogenides}

\subsection{Fe$_{1+y}$Te$_{1-x}$Se$_x$}

\begin{figure*}[t]
\includegraphics[width=15cm]{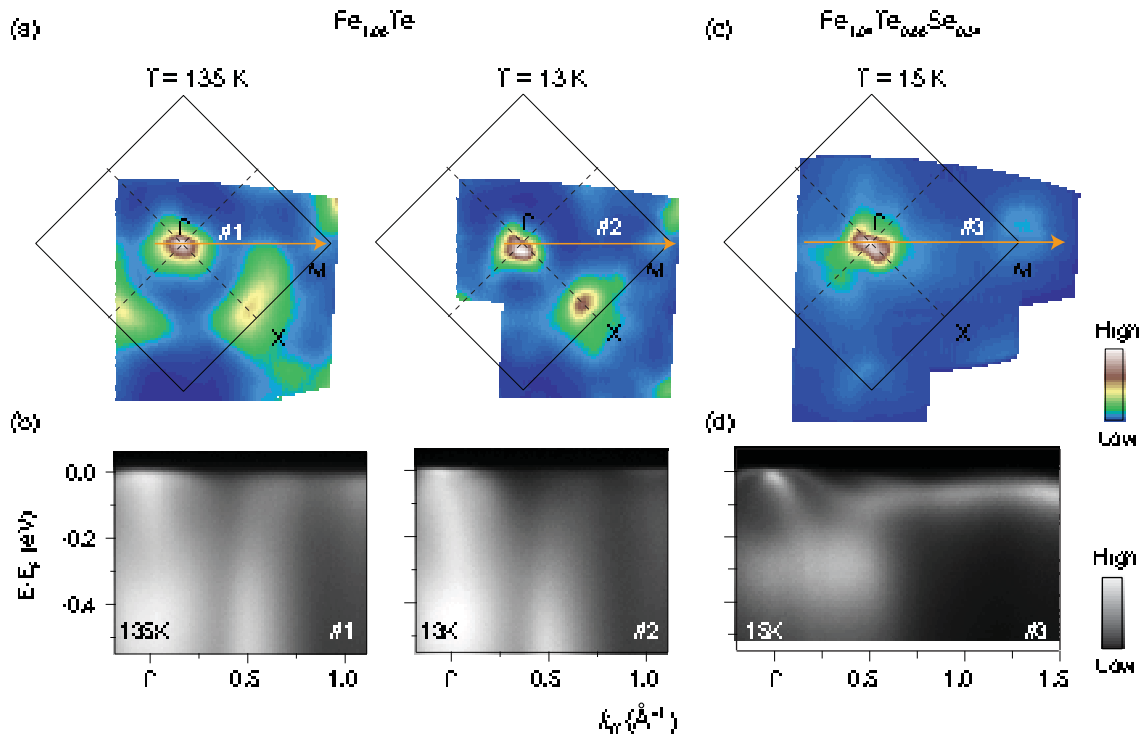}
\caption{(a) Fermi surface mapping, (b) photoemission intensity taken along the $\Gamma$-$M$ direction in Fe$_{1.06}$Te at 135~K and 13~K.  The data were taken with 24~eV photons. (c) Fermi surface mapping, (d) photoemission intensity taken along the $\Gamma$-$M$ direction in Fe$_{1.04}$Te$_{0.86}$Se$_{0.34}$ at 15~K . The data were taken with 22~eV photons.  The data on Fe$_{1.06}$Te is reprinted with permission from \cite{FeTePRB}, copyright 2010 by the American Physical Society. The data on Fe$_{1.04}$Te$_{0.86}$Se$_{0.34}$ is reprinted with permission from \cite{ChenPRB}, copyright 2010 by the American Physical Society.  } \label{FeTeSe}
\end{figure*}

The iron-pnictides and iron-chalcogenides have many propertities in common. The FeSe(Te) layer in Fe$_{1+y}$Te$_{1-x}$Se$_x$ is isostructural to the FeAs or FeP layer in iron pnictides. Fe$_{1+y}$Se shows superconductivity at a $T_C$ as high as 37~K under a 7~GPa hydrostatic pressure, which is comparable to the iron-pnictides. On the other hand, unlike the collinear SDW state in iron-pnictides, the magnetic ground state of Fe$_{1+y}$Te is a bicollinear commensurate or incommensurate antiferromagnetic state. Moreover, the magnetic moment in Fe$_{1+y}$Te is about 2 $\mu_B$, much lager than the 0.87 $\mu_B$ in BaFe$_2$As$_2$, or 0.36 $\mu_B$ in LaOFeAs.

The electronic structure of Fe$_{1.06}$Te is  shown in Figs.~\ref{FeTeSe}(a) and~\ref{FeTeSe}(b). The band structure is characterized by very broad features. The Fermi surface could not be clearly resolved. However, intensive spectral weight could be observed near the X point. Moreover, the spectral weight is redistributed significantly with decreasing temperature. Sharp coherent peaks could be observed near $E_F$ at 13~K  \cite{FeTePRB}. Considering all these results, we conclude that Fe$_{1.06}$Te distinguishes itself from iron-pnictides with much stronger local characters. The strong correlation effect and large spectral weight suppression near the $E_F$ could be responsible for the bicollinear magnetic order in Fe$_{1.06}$Te.

The superconducting Fe$_{1.04}$Te$_{0.66}$Se$_{0.34}$ with a $T_C$ of 15~K exhibits an electronic structure that resembles those of iron-pnictides [Fig.~\ref{FeTeSe}(c)] \cite{ChenPRB}.   There are three hole-like bands near the zone center and two electron-like bands near the zone corner. No spectral weight has been observed near the X point. The photoemission spectrum show sharp band dispersions, which is more coherent than that in Fe$_{1.06}$Te [Fig.~\ref{FeTeSe}(d)]. Therefore, the correlation effect is weakened and the Fermi surface topology changes significantly upon the Se doping, which explains the emergence of  superconductivity in Fe$_{1+y}$Te$_{1-x}$Se$_x$, and why it resembles  the iron-pnictides.


\subsection{Alkali-metal-intercalated iron selenide}

\subsubsection{Basic electronic structures of various phases and  phase seperation}

\begin{figure*}[t]
\includegraphics[width=16cm]{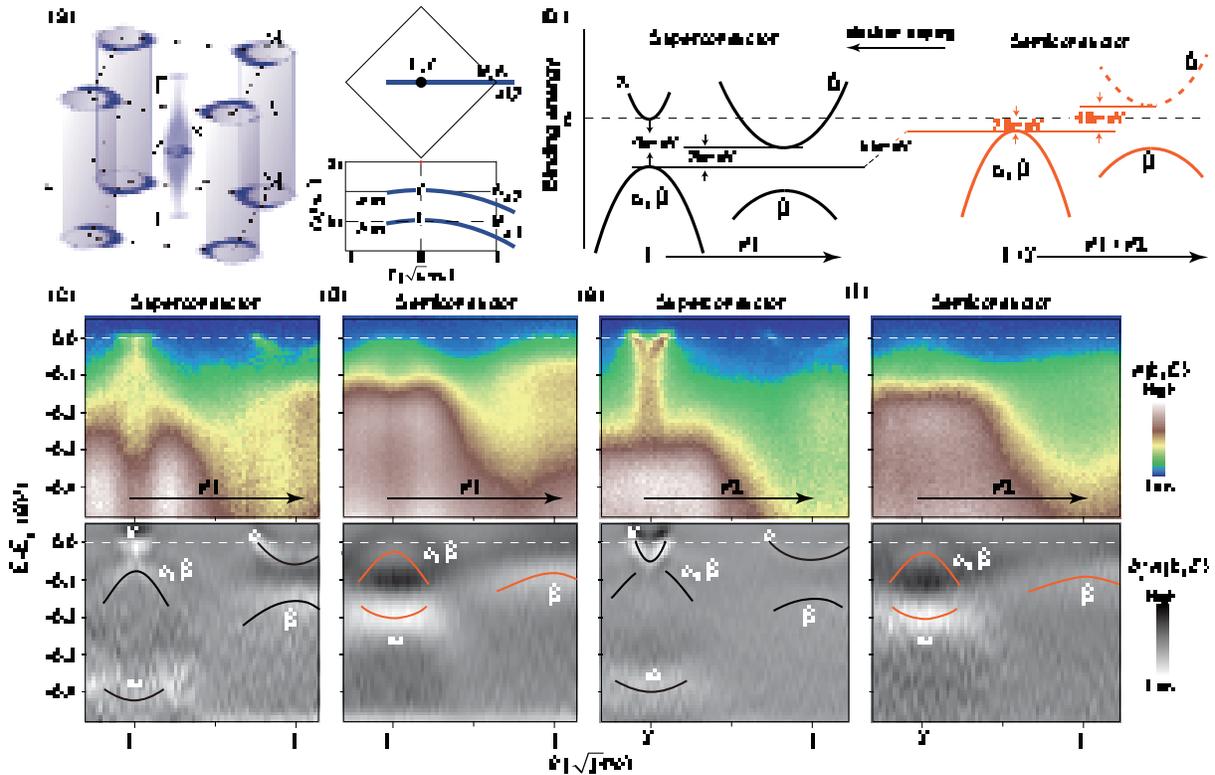}
\caption{Low energy electronic structures of the superconducting and
semiconducting  K$_x$Fe$_{2-y}$Se$_2$. (a) Left: the Fermi surface of the
superconducting phase in the 3D Brillouin zone. Right: the two momentum cut~\#1 and cut~\#2 sampled with 21~eV and  31~eV photons  respectively are illustrated in the two dimensional cross-sections of the 3D Brillouin zone. $c'$ is the distance between neighboring FeAs layers, and $a$ is the Fe-Fe distance in the FeAs plane. (b) The sketch of
the band structure evolution from the semiconductor to the
superconductor. (c) The photoemission intensities (upper panel) and
its second derivative with respect to energy (lower panel) along cut~\#1 across
$\Gamma$ for the superconductor taken at 35~K. (d)
is the same as (c) but for the semiconductor taken at 100~K. (e) and (f) are same as (c),
(d), except the data were taken along cut~\#2 across Z. Reprinted with permission from \cite{ChenPRX}, copyright 2011 by the American Physical Society. } \label{KFS1}
\end{figure*}

The discovery of A$_x$Fe$_{2-y}$Se$_2$ (A~=~K, Cs, Rb, ...) superconductor with $T_C$ of  about 31~K opens a new area in iron-based superconductor research \cite{kfsPRB}. The phase diagram is still under debate. Many phases have been discovered in A$_x$Fe$_{2-y}$Se$_2$, including several insulating phases with different magnetic and vacancy orders, semiconducting phase, and superconducting phase.   Therefore, one critical question is which phase is the parent phase for the superconductivity in A$_x$Fe$_{2-y}$Se$_2$ \cite{JQLi, QKXue, XHChen}.

Starting from the superconducting phase, the band structure is shown in Figs.~\ref{KFS1}(a), (c) and (e). In contrast to all the other iron-based superconductors, the Fermi surface of superconducting K$_x$Fe$_{2-y}$Se$_2$ only consists of electron pockets. There are two large electron pockets $\delta$/$\delta$' which are almost degenerated around each zone corner with little-$k_z$ dispersion and a small electron pocket $\kappa$ around the Z point. This rather unique electronic structure in K$_x$Fe$_{2-y}$Se$_2$ further highlights the diversity of the iron-based superconductors.

Compared with the superconducting phase, for semiconducting phase [Figs.~\ref{KFS1}(d) and~\ref{KFS1}(f)], the band top of the hole-like $\alpha$ and $\beta$ bands shifts up by 55~meV, and thus is about 20~meV below $E_F$. The electron pockets $\delta$/$\delta$' and $\kappa$ disappear near $E_F$ in the semiconducting phase. The electronic structures of the superconducting and semiconducting phases are summarized in Fig.~\ref{KFS1}(b). Because they have similar electronic structures, the semiconducting phase is perhaps a closer parent compound to the K$_x$Fe$_{2-y}$Se$_2$ superconductor than the insulators are. With electron doping, the semiconductor might evolve into a superconductor. Such an interesting semiconducting phase has largely been neglected in previous theoretical and experimental studies. It could be a possible starting point for modeling the superconductivity in K$_x$Fe$_{2-y}$Se$_2$, which is rather unique compared with other high-$T_C$ superconductors.

\begin{figure}[t]
\includegraphics[width=8.7cm]{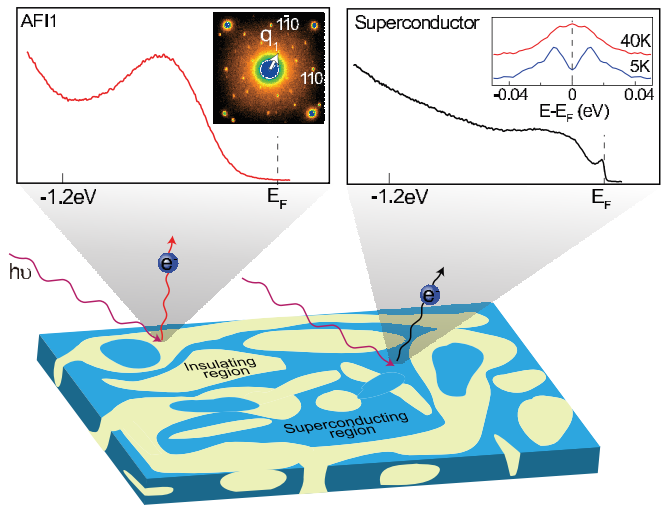}
\caption{ Cartoon for mesoscopic phase separation in
superconducting K$_x$Fe$_{2-y}$Se$_2$.  Different regions exhibit
different photoemission spectroscopic signature. Inset in the left
panel: the diffraction pattern of the $\sqrt{5} \times \sqrt{5}$
order was observed with TEM in both the superconductors and
semiconductors indicating a mixing of superconducting or
semiconducting phase with the AFI1 phase. The arrow in the
diffraction pattern indicates the superlattice modulation wave
vector $q_1$~=~(1/5,3/5,0). The TEM data was collected at room
temperature. Inset in the right panel: the symmetrized EDC's of the
superconductor across $T_c$, illustrating a superconducting gap. Reprinted with permission from \cite{ChenPRX}, copyright 2011 by the American Physical Society.  } \label{KFSphase}
\end{figure}

Based on the photoemission charging effects, we found that both the K$_x$Fe$_{2-y}$Se$_2$ superconductor and semicondtor are actually phase seprated \cite{ChenPRX}. As  sketched in Fig.~\ref{KFSphase}, the superconductor is consisted of an antiferromagnetic insulating phase (AFI1), and a superconducting phase. The AFI1 phase is characterized by the $\sqrt{5} \times \sqrt{5}$ vacancy order and a block antiferromagnetism as found in neutron scattering experiments. There is no density of state at the Fermi energy for AFI1.  Such a  phase separation in superconducting K$_x$Fe$_{2-y}$Se$_2$ is  mesoscopic, which has been further confirmed by transmission-electron-microscope (TEM) and STM measurements \cite{QKXue}.

\subsubsection{The superconducting gap distribution of K$_x$Fe$_{2-y}$Se$_2$ }

\begin{figure*}[t]
\includegraphics[width=16cm]{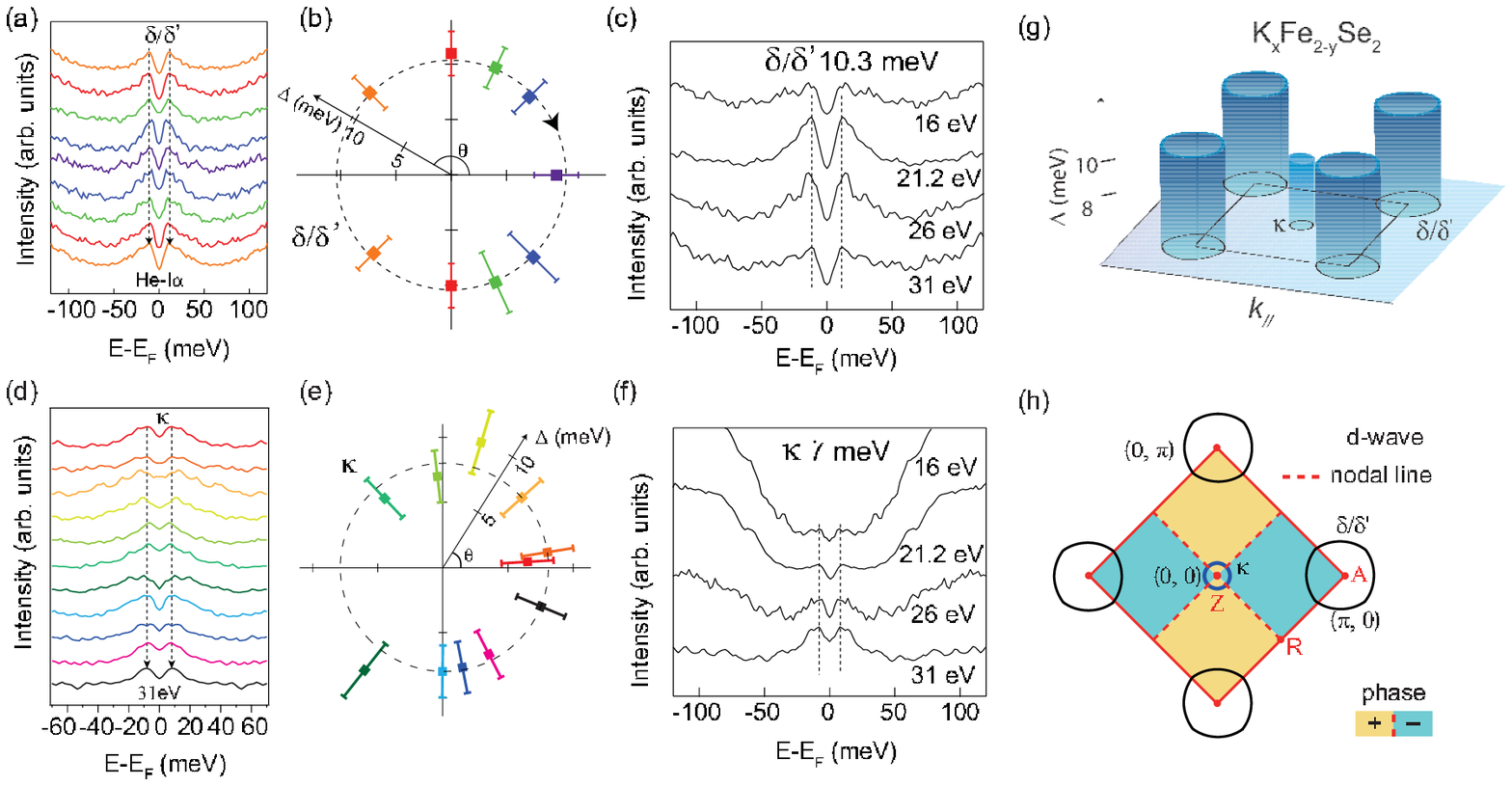}
\caption{(a) Symmetrized EDCs at various Fermi crossings for the $\delta$/$\delta$' electron pocket in superconducting K$_{x}$Fe$_{2-y}$Se$_2$. (b) Gap distribution of the $\delta$/$\delta$' electron pocket around M in polar coordinates, where the radius represents the gap, and the polar angle $\theta$ represents the position on the $\delta$/$\delta$' pocket with respect to M, with $\theta$~=~0 being the $M$-$\Gamma$ direction. (c) Photon energy dependence of the symmetrized EDCs for the $\delta$/$\delta$' electron pockets. All data were taken at 9~K. (d)$\sim$(f) are the same as (a)$\sim$(c), but for the $\kappa$ pocket. (g) Summary of the superconducting gap distribution at $\kappa$, $\delta$, and $\delta'$ pockets in K$_x$Fe$_{2-y}$Se$_2$. (h) Schematic diagram of the $d$-wave theoretical gap symmetry with the nodal lines along the $(0,0)-(\pm\pi,\pm\pi)$ directions. Positive and negative phases of the superconducting order parameter are denoted by different colors.  Panels (a)$\sim$(c) and (f) are reprinted with permission from \cite{KFeSe}, copyright 2011 by the Nature materials.  Panels (c), (e), (g) and (h) are reprinted with permission from \cite{Xu}, copyright 2012 by the American Physical Society.} \label{KFS3}
\end{figure*}

The spectral weight of insulating phases disappears near $E_F$ at low temperature due to the charging effect, so that the superconducting gap of the superconducting phase of K$_x$Fe$_{2-y}$Se$_2$ could be measured directly at low temperatures \cite{ChenPRX, KFeSe}. As shown in Fig.~\ref{KFS3}, the superconducting gap is isotropic with an amplitude of  $\sim$10.3~meV on the $\delta$/$\delta$' electron pockets [Figs.~\ref{KFS3}(a) and \ref{KFS3}(b)]. Further data taken with different photon energies in Fig.~\ref{KFS3}(c) indicate such a gap does not vary much with  $k_z$. The gap on the $\kappa$ pocket also shows isotropic character and little $k_z$-dependence, which is always about 7~meV [Figs.~\ref{KFS3}(d)$\sim$\ref{KFS3}(f)]. Figure~\ref{KFS3}(g) summarizes the superconducting gap distribution in the 3D momentum space.

The smaller gap at the zone center than those around the zone corner certainly violates the simple gap function of cos($k_x$)cos($k_y$) for the s$\pm$-pairing order parameter suggested for the iron pnictides \cite{KFeSe}. Moreover, as the Fermi surface size and spectral weight of the $\kappa$ band are minimal, its contribution to the superconductivity would be rather negligible with such a small gap. Therefore, the superconductivity in K$_x$Fe$_{2-y}$Se$_2$ should mainly rely on the large electron pockets around the zone corner. This thus rules out the common s${\pm}$ pairing symmetry proposal based on scattering between the hole and electron Fermi surface sheets in the iron-pnictides.

Theories based on local antiferromagnetic exchange interactions have predicted $s$-wave pairing symmetry in this system that can account for the experimental results \cite{Hu, Swave1}. However, calculations based on the scattering amongst the $\delta$/$\delta'$ electron pockets have indicated that the $d$-wave pairing channel would win over the $s$-wave pairing channel \cite{dwave1,dwave4}, which makes the superconducting order parameters to change sign between the neighboring  $\delta$/$\delta'$ Fermi pockets as illustrated in Fig.~\ref{KFS3}(h). Because  the four nodal $(0,0)-(\pm\pi,\pm\pi)$ directions (dashed lines)  do not cross any of the $\delta$/$\delta'$ Fermi cylinders, it is consistent with the observed nodeless gap structure on the $\delta$/$\delta'$ electron pockets \cite{KFeSe}. However as shown in Fig.~\ref{KFS3}(h), the nodal lines in the $d$-wave pairing scenario actually cross the $\kappa$  pocket, thus the observed isotropic superconducting gap on $\kappa$ [Figs.~\ref{KFS3}(d) and \ref{KFS3}(e)]  favors the $s$-wave pairing symmetry.  This helps pin down the pairing symmetry of K$_x$Fe$_{2-y}$Se$_2$.

\section{Summary}

Five years after the discovery of the superconductivity in iron-based superconductors, the mechanism of its superconductivity is yet to be revealed. However, ARPES has enabled us to unveil many critical information on iron-based superconductors and their parent compounds. We found that the low-lying electronic structures of iron-based superconductors are characterized by multi-band and multi-orbital nature. The Hund's rule coupling and the fluctuating collinear spin order is responsible for the large electronic structure reconstruction, spin density wave and structural transition in iron-pnictides. However, the magnetic-ordered state in Fe$_{1+y}$Te might be originated from the local antiferromagnetic exchange interactions. Upon doping, the superconductivity is enhanced by the Fermi surface nesting between $d_{xz}$/$d_{yz}$ orbitals while diminishes quickly after the central $d_{xz}$/$d_{yz}$ hole Fermi surfaces sink below $E_F$. The superconducting gap of both iron pnictides and iron selenides  might be described ubiquitously under the  $s$-wave pairing symmetry.  Particularly,  the  nodal ring near Z in  BaFe$_2$(As$_{1-x}$P$_x$)$_2$ is likely accidental, due to the strong mixing of the $d_{z^2}$ orbital on a hole Fermi surface sheet. The coexistence of SDW and superconductivity would cause anisotropic gap on the electron pockets. Moreover, the superconductivity  in K$_x$Fe$_{2-y}$Se$_2$ is dominated by the large electron pockets at the zone corner, leading to a reconsideration of the superconducting mechanism established in iron pnictides. These results lay the foundation for the ultimate understanding of unconventional high-$T_C$ superconductivity in iron-based superconductors.

Looking into the future, there are still many important issues that remain to be resovled. For example,  little is known on  the superconductivity of 48~K in (TlRb)$_{0.8}$Fe$_{1.67}$Se$_2$  under high pressure \cite{pressure},  and there are still arguments regarding  the nodal superconducting gap distribution in iron pnictides, thus more data need to be accumlated on
 materials like the overdoped BaFe$_2$(As$_{1-x}$P$_x$)$_2$, BaFe$_{2-x}$Ru$_x$As$_2$ and so on.  Moreover, a novel and  complex  pairing symmetry for the iron based superconductors have been proposed by Hu et al. \cite{Hupairing} which are of the utter importance to be settled experimentally. On the other hand,  intriguing novel pheonomena  continuously emerge. Signs for superconductivity has been found at 65~K in mono-layer FeSe thin film on STO substrate \cite{zhou, tan}, which suggests that there is still room for futher enhancement of $T_C$. Therefore, there are still plenty to be explored, and as always  in this field,  more surprises and discoveries are expected.

Acknowledgements: We are grateful for many collaborators for providing us the materials,  helpful discussions, and experimental support at various synchrotron  beamlines in the last five years. This work is supported in part by the National Science Foundation of China and National Basic Research Program of China (973 Program) under the grant Nos. 2012CB921400,
2011CB921802, 2011CBA00112.

\end{document}